\documentclass[11pt]{article}
\usepackage[dvips]{graphicx}
\usepackage{epsfig}
\usepackage{amssymb}
\usepackage{color}
\usepackage{amsmath}
\usepackage{nicefrac}
\usepackage[left]{lineno}
\usepackage{hyperref}
\usepackage{adjustbox}
\usepackage{xspace}
\usepackage{multirow}

\definecolor{burgundy}{rgb}{0.5, 0.0, 0.13}

\def\geant {\mbox{\textsc{Geant4}}\xspace}

\begin{document}
\centerline{\LARGE EUROPEAN ORGANIZATION FOR NUCLEAR RESEARCH}

%
\vspace{10mm} {\flushright{
CERN-EP-2019-104 \\
18 May 2019\\
\vspace{4mm}
Revised version:\\30 June 2019\\
}}
\vspace{-30mm}

%
%

%
\vspace{40mm}

\begin{center}
\boldmath
{\bf {\Large\boldmath{Searches for lepton number violating $K^+$ decays}}}
\unboldmath
\end{center}
\vspace{4mm}
\begin{center}
{\Large The NA62 Collaboration}\\
\end{center}

\begin{abstract}
The NA62 experiment at CERN reports a search for the lepton number violating decays  $K^+\to\pi^-e^+e^+$ and $K^+\to\pi^-\mu^+\mu^+$ using a data sample collected in 2017. No signals are observed, and upper limits on the branching fractions of these decays of $2.2\times 10^{-10}$ and $4.2\times 10^{-11}$ are obtained, respectively, at 90\% confidence level. These upper limits improve on previously reported measurements by factors of 3 and 2, respectively.
\end{abstract}

\begin{center}
{\it Accepted for publication in Physics Letters B}
\end{center}

\newpage
\begin{center}
{\Large The NA62 Collaboration$\,$\renewcommand{\thefootnote}{\fnsymbol{footnote}}%
\footnotemark[1]\renewcommand{\thefootnote}{\arabic{footnote}}}\\
\end{center}
\vspace{2mm}
\begin{raggedright}
\noindent
{\bf Universit\'e Catholique de Louvain, Louvain-La-Neuve, Belgium}\\
 E.~Cortina Gil,
 A.~Kleimenova,
 E.~Minucci$\,$\footnotemark[1],
 S.~Padolski$\,$\footnotemark[2],
 P.~Petrov,
 A.~Shaikhiev$\,$\footnotemark[3],
 R.~Volpe\\[2mm]

{\bf TRIUMF, Vancouver, British Columbia, Canada}\\
 T.~Numao,
 Y.~Petrov,
 B.~Velghe\\[2mm]

{\bf University of British Columbia, Vancouver, British Columbia, Canada}\\
 D.~Bryman,
 J.~Fu$\,$\footnotemark[4]\\[2mm]

{\bf Charles University, Prague, Czech Republic}\\
 T.~Husek$\,$\footnotemark[5],
 J.~Jerhot,
 K.~Kampf,
 M.~Zamkovsky\\[2mm]

{\bf Institut f\"ur Physik and PRISMA Cluster of excellence, Universit\"at Mainz, Mainz, Germany}\\
 R.~Aliberti,
 G.~Khoriauli$\,$\footnotemark[6],
 J.~Kunze,
 D.~Lomidze$\,$\footnotemark[7],
 L.~Peruzzo,
 M.~Vormstein,
 R.~Wanke\\[2mm]

{\bf Dipartimento di Fisica e Scienze della Terra dell'Universit\`a e INFN, Sezione di Ferrara, Ferrara, Italy}\\
 P.~Dalpiaz,
 M.~Fiorini,
 I.~Neri,
 A.~Norton,
 F.~Petrucci,
 H.~Wahl\\[2mm]

{\bf INFN, Sezione di Ferrara, Ferrara, Italy}\\
 A.~Cotta Ramusino,
 A.~Gianoli\\[2mm]

{\bf Dipartimento di Fisica e Astronomia dell'Universit\`a e INFN, Sezione di Firenze, Sesto Fiorentino, Italy}\\
 E.~Iacopini,
 G.~Latino,
 M.~Lenti,
 A.~Parenti\\[2mm]

{\bf INFN, Sezione di Firenze, Sesto Fiorentino, Italy}\\
 A.~Bizzeti$\,$\footnotemark[8],
 F.~Bucci\\[2mm]

{\bf Laboratori Nazionali di Frascati, Frascati, Italy}\\
 A.~Antonelli,
 G.~Georgiev$\,$\footnotemark[9],
 V.~Kozhuharov$\,$\footnotemark[9],
 G.~Lanfranchi,
 G.~Mannocchi,
 S.~Martellotti,
 M.~Moulson,
 T.~Spadaro\\[2mm]

{\bf Dipartimento di Fisica ``Ettore Pancini'' e INFN, Sezione di Napoli, Napoli, Italy}\\
 F.~Ambrosino,
 T.~Capussela,
 M.~Corvino,
 D.~Di Filippo,
 P.~Massarotti,
 M.~Mirra,
 M.~Napolitano,
 G.~Saracino\\[2mm]

{\bf Dipartimento di Fisica e Geologia dell'Universit\`a e INFN, Sezione di Perugia, Perugia, Italy}\\
 G.~Anzivino,
 F.~Brizioli,
 E.~Imbergamo,
 R.~Lollini,
 R.~Piandani,
 C.~Santoni\\[2mm]

{\bf INFN, Sezione di Perugia, Perugia, Italy}\\
 M.~Barbanera$\,$\footnotemark[10],
 P.~Cenci,
 B.~Checcucci,
 P.~Lubrano,
 M.~Lupi$\,$\footnotemark[11],
 M.~Pepe,
 M.~Piccini\\[2mm]

{\bf Dipartimento di Fisica dell'Universit\`a e INFN, Sezione di Pisa, Pisa, Italy}\\
 F.~Costantini,
 L.~Di Lella,
 N.~Doble,
 M.~Giorgi,
 S.~Giudici,
 G.~Lamanna,
 E.~Lari,
 E.~Pedreschi,
 M.~Sozzi\\[2mm]

{\bf INFN, Sezione di Pisa, Pisa, Italy}\\
 C.~Cerri,
 R.~Fantechi,
 L.~Pontisso,
 F.~Spinella\\[2mm]

{\bf Scuola Normale Superiore e INFN, Sezione di Pisa, Pisa, Italy}\\
 I.~Mannelli\\[2mm]
\clearpage
{\bf Dipartimento di Fisica, Sapienza Universit\`a di Roma e INFN, Sezione di Roma I, Roma, Italy}\\
 G.~D'Agostini,
 M.~Raggi\\[2mm]

{\bf INFN, Sezione di Roma I, Roma, Italy}\\
 A.~Biagioni,
 E.~Leonardi,
 A.~Lonardo,
 P.~Valente,
 P.~Vicini\\[2mm]

{\bf INFN, Sezione di Roma Tor Vergata, Roma, Italy}\\
 R.~Ammendola,
 V.~Bonaiuto$\,$\footnotemark[12],
 A.~Fucci,
 A.~Salamon,
 F.~Sargeni$\,$\footnotemark[13]\\[2mm]

{\bf Dipartimento di Fisica dell'Universit\`a e INFN, Sezione di Torino, Torino, Italy}\\
 R.~Arcidiacono$\,$\footnotemark[14],
 B.~Bloch-Devaux,
 M.~Boretto$\,$\footnotemark[15],
 E.~Menichetti,
 E.~Migliore,
 D.~Soldi\\[2mm]

{\bf INFN, Sezione di Torino, Torino, Italy}\\
 C.~Biino,
 A.~Filippi,
 F.~Marchetto\\[2mm]

{\bf Instituto de F\'isica, Universidad Aut\'onoma de San Luis Potos\'i, San Luis Potos\'i, Mexico}\\
 J.~Engelfried,
 N.~Estrada-Tristan$\,$\footnotemark[16]\\[2mm]

{\bf Horia Hulubei national Institute of Physics and Nuclear Engineering, Bucharest-Magurele, Romania}\\
 A. M.~Bragadireanu,
 S. A.~Ghinescu,
 O. E.~Hutanu\\[2mm]

{\bf Joint Institute for Nuclear Research, Dubna, Russia}\\
 T.~Enik,
 V.~Falaleev,
 V.~Kekelidze,
 A.~Korotkova,
 L.~Litov$\,$\footnotemark[9],
 D.~Madigozhin,
 M.~Misheva$\,$\footnotemark[17],
 N.~Molokanova,
 S.~Movchan,
 I.~Polenkevich,
 Yu.~Potrebenikov,
 S.~Shkarovskiy,
 A.~Zinchenko$\,$\renewcommand{\thefootnote}{\fnsymbol{footnote}}\footnotemark[2]\renewcommand{\thefootnote}{\arabic{footnote}}\\[2mm]

{\bf Institute for Nuclear Research of the Russian Academy of Sciences, Moscow, Russia}\\
 S.~Fedotov,
 E.~Gushchin,
 A.~Khotyantsev,
 Y.~Kudenko$\,$\footnotemark[18],
 V.~Kurochka,
 M.~Medvedeva,
 A.~Mefodev\\[2mm]

{\bf Institute for High Energy Physics - State Research Center of Russian Federation, Protvino, Russia}\\
 S.~Kholodenko,
 V.~Kurshetsov,
 V.~Obraztsov,
 A.~Ostankov,
 V.~Semenov$\,$\renewcommand{\thefootnote}{\fnsymbol{footnote}}\footnotemark[2]\renewcommand{\thefootnote}{\arabic{footnote}},
 V.~Sugonyaev,
 O.~Yushchenko\\[2mm]

{\bf Faculty of Mathematics, Physics and Informatics, Comenius University, Bratislava, Slovakia}\\
 L.~Bician,
 T.~Blazek,
 V.~Cerny,
 Z.~Kucerova\\[2mm]

{\bf CERN,  European Organization for Nuclear Research, Geneva, Switzerland}\\
 J.~Bernhard,
 A.~Ceccucci,
 H.~Danielsson,
 N.~De Simone$\,$\footnotemark[19],
 F.~Duval,
 B.~D\"obrich,
 L.~Federici,
 E.~Gamberini,
 L.~Gatignon,
 R.~Guida,
 F.~Hahn$\,$\renewcommand{\thefootnote}{\fnsymbol{footnote}}\footnotemark[2]\renewcommand{\thefootnote}{\arabic{footnote}},
 E. B.~Holzer,
 B.~Jenninger,
 M.~Koval,
 P.~Laycock$\,$\footnotemark[2],
 G.~Lehmann Miotto,
 P.~Lichard,
 A.~Mapelli,
 R.~Marchevski,
 K.~Massri,
 M.~Noy,
 V.~Palladino$\,$\footnotemark[20],
 M.~Perrin-Terrin$\,$\footnotemark[21]$^,\,$\footnotemark[22],
 J.~Pinzino,
 V.~Ryjov,
 S.~Schuchmann,
 S.~Venditti\\[2mm]

{\bf University of Birmingham, Birmingham, United Kingdom}\\
 T.~Bache,
 M. B.~Brunetti,
 V.~Duk,
 V.~Fascianelli$\,$\footnotemark[23],
 J. R.~Fry,
 F.~Gonnella,
 E.~Goudzovski\renewcommand{\thefootnote}{\fnsymbol{footnote}}%
\footnotemark[1]\renewcommand{\thefootnote}{\arabic{footnote}},
 L.~Iacobuzio,
 C.~Lazzeroni,
 N.~Lurkin,
 F.~Newson,
 C.~Parkinson,
 A.~Romano,
 A.~Sergi,
 A.~Sturgess,
 J.~Swallow\\[2mm]

{\bf University of Bristol, Bristol, United Kingdom}\\
 H.~Heath,
 R.~Page,
 S.~Trilov\\[2mm]

{\bf University of Glasgow, Glasgow, United Kingdom}\\
 B.~Angelucci,
 D.~Britton,
 C.~Graham,
 D.~Protopopescu\\[2mm]

{\bf University of Lancaster, Lancaster, United Kingdom}\\
 J.~Carmignani,
 J. B.~Dainton,
 R. W. L.~Jones,
 G.~Ruggiero\\[2mm]

{\bf University of Liverpool, Liverpool, United Kingdom}\\
 L.~Fulton,
 D.~Hutchcroft,
 E.~Maurice$\,$\footnotemark[24],
 B.~Wrona\\[2mm]

{\bf George Mason University, Fairfax, Virginia, USA}\\
 A.~Conovaloff,
 P.~Cooper,
 D.~Coward$\,$\footnotemark[25],
 P.~Rubin\\[2mm]

\end{raggedright}
%
%
\setcounter{footnote}{0}
\renewcommand{\thefootnote}{\fnsymbol{footnote}}
\footnotetext[1]{Corresponding author: Evgueni Goudzovski. Email: Evgueni.Goudzovski@cern.ch}
\footnotetext[2]{Deceased}
\renewcommand{\thefootnote}{\arabic{footnote}}

\footnotetext[1]{Present address: Laboratori Nazionali di Frascati, I-00044 Frascati, Italy}
\footnotetext[2]{Present address: Brookhaven National Laboratory, Upton, NY 11973, USA}
\footnotetext[3]{Also at Institute for Nuclear Research of the Russian Academy of Sciences, 117312 Moscow, Russia}
\footnotetext[4]{Present address: UCLA Physics and Biology in Medicine, Los Angeles, CA 90095, USA}
\footnotetext[5]{Present address: IFIC, Universitat de Val\`encia - CSIC, E-46071 Val\`encia, Spain}
\footnotetext[6]{Present address: Universit\"at W\"urzburg, D-97070 W\"urzburg, Germany}
\footnotetext[7]{Present address: Universit\"at Hamburg, D-20146 Hamburg, Germany}
\footnotetext[8]{Also at Dipartimento di Fisica, Universit\`a di Modena e Reggio Emilia, I-41125 Modena, Italy}
\footnotetext[9]{Also at Faculty of Physics, University of Sofia, BG-1164 Sofia, Bulgaria}
\footnotetext[10]{Present address: INFN, Sezione di Pisa, I-56100 Pisa, Italy}
\footnotetext[11]{Present address: Institut am Fachbereich Informatik und Mathematik, Goethe Universit\"at, D-60323 Frankfurt am Main, Germany}
\footnotetext[12]{Also at Department of Industrial Engineering, University of Roma Tor Vergata, I-00173 Roma, Italy}
\footnotetext[13]{Also at Department of Electronic Engineering, University of Roma Tor Vergata, I-00173 Roma, Italy}
\footnotetext[14]{Also at Universit\`a degli Studi del Piemonte Orientale, I-13100 Vercelli, Italy}
\footnotetext[15]{Also at CERN,  European Organization for Nuclear Research, CH-1211 Geneva 23, Switzerland}
\footnotetext[16]{Also at Universidad de Guanajuato, Guanajuato, Mexico}
\footnotetext[17]{Present address: Institute of Nuclear Research and Nuclear Energy of Bulgarian Academy of Science (INRNE-BAS), BG-1784 Sofia, Bulgaria}
\footnotetext[18]{Also at National Research Nuclear University (MEPhI), 115409 Moscow and Moscow Institute of Physics and Technology, 141701 Moscow region, Moscow, Russia}
\footnotetext[19]{Present address: DESY, D-15738 Zeuthen, Germany}
\footnotetext[20]{Present address: Physics Department, Imperial College London, London, SW7 2BW, UK}
\footnotetext[21]{Present address: Centre de Physique des Particules de Marseille, Universit\'e Aix Marseille, CNRS/IN2P3, F-13288, Marseille, France}
\footnotetext[22]{Also at Universit\'e Catholique de Louvain, B-1348 Louvain-La-Neuve, Belgium}
\footnotetext[23]{Present address: Dipartimento di Psicologia, Universit\`a di Roma La Sapienza, I-00185 Roma, Italy}
\footnotetext[24]{Present address: Laboratoire Leprince Ringuet, F-91120 Palaiseau, France}
\footnotetext[25]{Also at SLAC National Accelerator Laboratory, Stanford University, Menlo Park, CA 94025, USA}

\newpage


\section*{Introduction}

In the Standard Model (SM), neutrinos are strictly massless due to the absence of right-handed chiral states. The discovery of neutrino oscillations has conclusively demonstrated that neutrinos have non-zero masses. Therefore the observation of lepton number violating processes involving charged leptons would verify the Majorana nature of the neutrino.

The decays of the charged kaon $K^+\to\pi^-\ell^+\ell^+$ (where $\ell=e,\mu$), violating conservation of lepton number by two units, may be mediated by a massive Majorana neutrino~\cite{li00,at09}.
The current limits at 90\% CL on the branching fractions of these decays are ${\cal B}(K^+\to\pi^-e^+e^+)<6.4\times 10^{-10}$ obtained by the BNL E865 experiment~\cite{ap00}, and ${\cal B}(K^+\to\pi^-\mu^+\mu^+)<8.6\times 10^{-11}$ obtained by the CERN NA48/2 experiment~\cite{ba17}. A search for these processes in about 30\% of the data collected by the NA62 experiment at CERN in 2016--18 is reported here.


\section{Beam, detector and data sample}
\label{sec:detector}

\begin{figure}[t]
\begin{center}
\resizebox{\textwidth}{!}{\includegraphics{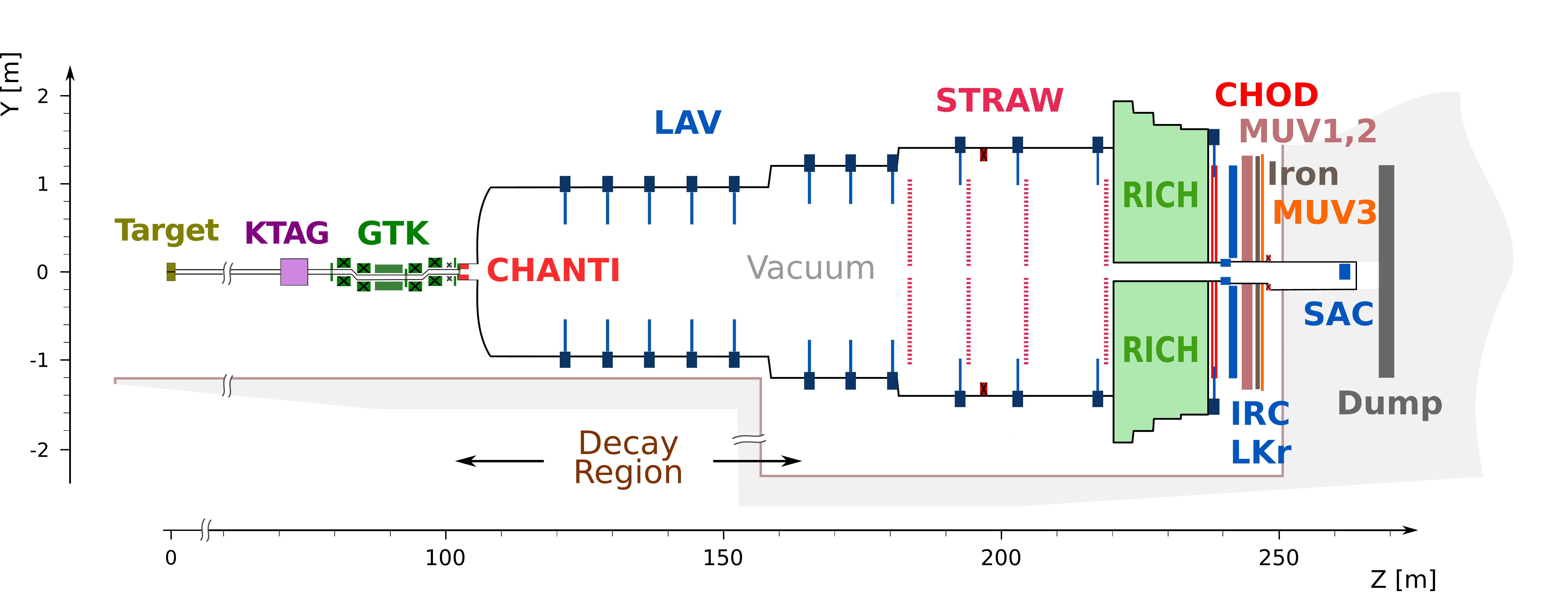}}
\put(-361,64){\tiny\color{burgundy}\rotatebox{90}{\bf GTK1}}
\put(-340,64){\tiny\color{burgundy}\rotatebox{90}{\bf GTK2}}
\put(-327,100){\tiny\color{burgundy}\rotatebox{90}{\bf TRIM5}}
\put(-325,64){\tiny\color{burgundy}\rotatebox{90}{\bf GTK3}}
\put(-188,35){\tiny\color{burgundy}\rotatebox{40}{\bf MNP33}}
\end{center}
\vspace{-15mm}
\caption{Schematic side view of the NA62 beamline and detector.}
\label{fig:detector}
\end{figure}
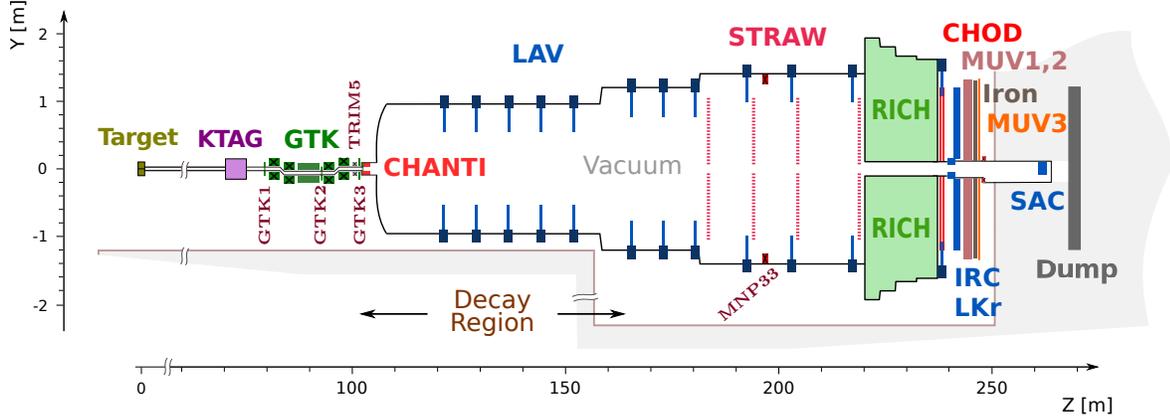

The layout of the NA62 beamline and detector~\cite{na62-detector} is shown schematically in Fig.~\ref{fig:detector}. An unseparated beam of $\pi^+$ (70\%), protons (23\%) and $K^+$ (6\%) is created by directing 400~GeV/$c$ protons extracted from the CERN SPS onto a beryllium target in spills of 3~s effective duration. The nominal central momentum of this secondary beam is 75~GeV/$c$ with a momentum spread of 1\% (rms).
Beam kaons are tagged with 70~ps time resolution by a differential Cherenkov counter (KTAG) using a nitrogen radiator at 1.75~bar pressure contained in a 5~m long vessel. Beam particle momenta are measured by a three-station silicon pixel spectrometer (GTK); inelastic interactions of beam particles with the last station (GTK3) are detected by an array of scintillator hodoscopes (CHANTI). A dipole magnet (TRIM5) providing a 90~MeV/$c$ horizontal momentum kick is located in front of GTK3. The beam is delivered into a vacuum tank containing a 75~m long fiducial decay volume (FV) starting 2.6~m downstream of GTK3. The beam divergence at the FV entrance is 0.11~mrad (rms) in both horizontal and vertical planes. Downstream of the FV, undecayed beam particles continue their path in vacuum.

Momenta of charged particles produced in $K^+$ decays in the FV are measured by a magnetic spectrometer (STRAW) located in the vacuum tank downstream of the FV. The spectrometer consists of four tracking chambers made of straw tubes, and a dipole magnet (MNP33) located between the second and third chambers providing a horizontal momentum kick of 270~MeV/$c$ in a direction opposite to that produced by TRIM5. The achieved momentum resolution $\sigma_p/p$ lies in the range of 0.3--0.4\%.

A ring-imaging Cherenkov detector (RICH), consisting of a 17.5~m long vessel filled with neon at atmospheric pressure, is used for the identification and time measurement of charged particles. The RICH provides a reference trigger time, typically with 70~ps precision. The Cherenkov threshold for pions is 12.5~GeV/$c$. Positively and negatively charged particles have different angular distributions downstream of the MNP33 magnet; the RICH optical system is optimized to collect light emitted by positively charged particles. Two scintillator hodoscopes
CHOD, which include a matrix of tiles, as well as two orthogonal planes of slabs, arranged in four quadrants)
downstream of the RICH provide trigger signals and time measurements with 200~ps precision.

A $27X_0$ thick quasi-homogeneous liquid krypton (LKr) electromagnetic calorimeter is used for particle identification and photon detection. The calorimeter has an active volume of 7~m$^3$, is segmented in the transverse direction into 13248 projective cells of approximately $2\!\times\!2$~cm$^2$, and provides an energy resolution of $\sigma_E/E=(4.8/\sqrt{E}\oplus11/E\oplus0.9)\%$, where $E$ is expressed in GeV. To achieve hermetic acceptance for photons emitted in $K^+$ decays in the FV at angles up to 50~mrad to the beam axis, the LKr calorimeter is supplemented by annular lead glass detectors (LAV) installed in 12~positions around and downstream of the FV, and two lead/scintillator sampling calorimeters (IRC, SAC) located close to the beam axis. An iron/scintillator sampling hadronic calorimeter formed of two modules (MUV1,2) and a muon detector (MUV3) consisting of 148~scintillator tiles located behind an 80~cm thick iron wall are used for particle identification.


The data sample used for this analysis is obtained from $2.3\times 10^5$ SPS spills recorded over three months of operation in 2017. The typical beam intensity was \mbox{$2.0\times 10^{12}$} protons per spill, corresponding to a mean instantaneous beam particle rate at the FV entrance of 450~MHz, and a mean $K^+$ decay rate in the FV of 3.5~MHz. Dedicated multi-track, di-electron and di-muon trigger chains are used. The low-level (L0) multi-track trigger is based on RICH signal multiplicity and a requirement for a coincidence of signals in two opposite CHOD quadrants. The di-electron L0 trigger additionally requires that at least 20 GeV of energy is deposited in the LKr calorimeter, while the di-muon L0 trigger requires a coincidence of signals from two MUV3 tiles. The software (L1) trigger involves beam $K^+$ identification by KTAG and reconstruction of a negatively charged track in STRAW. For signal-like samples,
the measured inefficiencies of the CHOD (STRAW) conditions are at the 2\%~(4\%) level, while those of the other trigger components are of the order of $10^{-3}$. The multi-track, di-electron and di-muon trigger chains were downscaled typically by factors of 100, 8 and 2, respectively.


\section{Event selection}
\label{sec:selection}

The processes of interest $K^+\to\pi^-\ell^+\ell^+$ (denoted ``LNV decays'') and the flavour-changing neutral current decays $K^+\to\pi^+\ell^+\ell^-$ (denoted ``SM decays'') are
collected concurrently through the same trigger chains. The SM decays with ${\cal O}(10^{-7})$ branching fractions known experimentally to a few percent accuracy~\cite{pdg} are used for normalization. Under the assumption of similar kinematic distributions, this approach leads to first-order cancellation
of the effects of detector inefficiencies, trigger inefficiencies and pileup. Both the LNV and SM decays with electrons (muons) in the final state are denoted as $K_{\pi ee}$ ($K_{\pi\mu\mu}$), and collectively as $K_{\pi\ell\ell}$. The principal selection criteria for $K_{\pi\ell\ell}$ decays are listed below.
\begin{itemize}
\item The di-electron and multi-track trigger chains are used to collect $K_{\pi ee}$ candidates, and the di-muon trigger chain is used to collect $K_{\pi\mu\mu}$ candidates.
\item Three-track vertices are reconstructed by extrapolation of STRAW tracks upstream into the FV, taking into account the measured residual magnetic field in the vacuum tank, and selecting triplets of tracks consistent with originating from the same point. The presence of exactly one vertex is required. The vertex should be located within the FV and have a total electric charge of $q=+1$. The extrapolation of the selected tracks into the transverse planes of the downstream detectors should be within the corresponding geometrical acceptance. Each pair of selected tracks should be separated by at least 15~mm in the first STRAW chamber plane to suppress photon conversions and fake tracks, and in the $K_{\pi ee}$ case by at least 200~mm in the LKr front plane to avoid shower overlap.
\item Reconstructed track momenta should be $8~(5)~{\rm GeV}/c<p<45~{\rm GeV}/c$ in the $K_{\pi ee}$ ($K_{\pi\mu\mu}$) case. The total momentum, $p_{\rm vtx}$, of the three tracks should satisfy the condition $|p_{\rm vtx}-p_{\rm beam}|<2.5~{\rm GeV}/c$, where $p_{\rm beam}$ is the central beam momentum. The total transverse momentum with respect to the beam axis should be $p_T<30~{\rm MeV}/c$. The quantity $p_{\rm beam}$ and the beam axis direction are measured continuously using fully reconstructed $K^+\to\pi^+\pi^+\pi^-$ decays.
\item Track times are defined using CHOD information, as well as RICH information in the $K_{\pi ee}$ case. The vertex tracks are required to be in time within 15~ns of each other. The vertex time is defined as a weighted average of the track times, taking into account CHOD and RICH time resolution.
\item Pion candidates are required to have the ratio of energy deposition in the LKr calorimeter to momentum measured by the spectrometer $E/p<0.85~(0.9)$ in the $K_{\pi ee}$ ($K_{\pi\mu\mu}$) case, and no associated in-time MUV3 signals in the $K_{\pi\mu\mu}$ case.  Electron ($e^\pm$) candidates are required to have $0.9<E/p<1.1$. Muon candidates are identified by requiring $E/p<0.2$ and a geometrically associated MUV3 signal within 5~ns of the vertex time. The vertex should include a pion candidate and two lepton candidates of the same flavour. The conditions used for $\pi^\pm$, $e^\pm$ and $\mu^\pm$ identification are mutually exclusive within each selection.
\end{itemize}
The following additional conditions are applied in the $K_{\pi ee}$ case.
\begin{itemize}
\item An identification algorithm based on the likelihoods of mass hypotheses evaluated using the RICH signal pattern~\cite{mu94} is applied to $e^+$ candidates. The algorithm considers each track independently. The angles between track pairs in the RICH are required to exceed~4 mrad to reduce overlaps between Cherenkov light-cones, decreasing the acceptance of both the SM and LNV selections by 7\% in relative terms. A selection without $e^+$ identification in the RICH and without the angular separation requirement is used for background validation; it is referred to as the {\it auxiliary selection}, as opposed to the {\it standard selection}.
\item To suppress backgrounds from $K^+\to\pi^+\pi^0_D$ and $K^+\to\pi^0_D e^+\nu$ decays followed by the $\pi^0_D\to e^+e^-\gamma$ decay, which are characterized by emission of soft photons at large angles, no signals are allowed in the LAV detectors within 4~ns of the vertex time. Photon veto conditions in the LKr, IRC and SAC calorimeters are not applied, as the background events with energetic photons emitted forward are suppressed by the momentum ($p_{\rm vtx}$) condition.
\item For the SM decay, a requirement on the reconstructed $e^+e^-$ mass $m_{ee}>140~{\rm MeV}/c^2$ is applied to suppress backgrounds from the $K^+\to\pi^+\pi^0$ decay followed by $\pi^0_D\to e^+e^-\gamma$, $\pi^0_{DD}\to e^+e^-e^+e^-$ and $\pi^0\to e^+e^-$ decays.\footnote{It should be noted however that the $K^+\to\pi^+e^+e^-$ decay is observed with negligible background also in the mass range $m_{ee}<100~{\rm MeV}/c^2$.} This leads to a 27\% reduction of acceptance in relative terms. For the LNV decay, these backgrounds contribute only via double particle misidentification, and kinematic suppression is therefore not required.
\end{itemize}

For the SM decays, the signal regions are defined in terms of the reconstructed $\pi\ell\ell$ mass as $470~{\rm MeV}/c^2<m_{\pi ee}<505~{\rm MeV}/c^2$ in the $K_{\pi ee}$ case (asymmetric with respect to the nominal $K^+$ mass $m_K$~\cite{pdg} to account for the radiative tail), and $484~{\rm MeV}/c^2<m_{\pi\mu\mu}<504~{\rm MeV}/c^2$ in the $K_{\pi\mu\mu}$ case. For LNV decays, the mass regions defined above were masked for data events until the completion of the background evaluation. The LNV signal mass regions are defined by tighter conditions $|m_{\pi\ell\ell}-m_K|<3\cdot\delta m_{\pi\ell\ell}$, where $\delta m_{\pi ee}=1.7~{\rm MeV}/c^2$ and $\delta m_{\pi\mu\mu}=1.1~{\rm MeV}/c^2$ are the mass resolutions measured from the data for the SM decays. The control regions $m_{\pi ee}<470~{\rm MeV}/c^2$ and $m_{\pi\mu\mu}<484~{\rm MeV}/c^2$ within both the SM and LNV selections were used for validation of the background evaluation procedures.


\section{Background evaluation}
\label{sec:bkg}

Acceptances and backgrounds are evaluated using Monte Carlo (MC) simulation based on the \geant toolkit~\cite{geant4} to describe detector geometry and response. Certain aspects of the simulation are tuned using input from the data, and data-driven methods are employed to address specific background sources.

\boldmath
\subsection{$K_{\pi ee}$ analysis}
\label{sec:piee}
\unboldmath

Backgrounds to the $K_{\pi ee}$ processes arise from misidentification of pions as electrons and vice versa. Background evaluation is based on simulations involving the measured pion ($\pi^\pm$) and electron ($e^\pm$) identification efficiencies $\varepsilon_{\pi}^\pm$, $\varepsilon_{e}^\pm$, as well as pion to electron ($P_{\pi e}^\pm$) and electron to pion ($P_{e\pi}^\pm$) misidentification probabilities. Each quantity is measured as a function of momentum using pion and positron samples obtained from kinematic selections of $K^+\to\pi^+\pi^+\pi^-$ and $K^+\to\pi^0e^+\nu$ decays, with the residual $K^+\to\pi^+\pi^0$ background subtracted in the latter case. The results of the measurements are summarized in Table~\ref{tab:pid}. The LKr calorimeter response is known to be the same for electrons and positrons~\cite{na48}. The typical inefficiencies $1-\varepsilon_{\pi,e}^\pm$ and misidentification probabilities are ${\cal O}(10^{-2})$ with weak momentum dependence, except for the $\pi^+$ misidentification probability $P_{\pi e}^+$ which has a minimum of $10^{-5}$ at a momentum of 25~GeV/$c$, and increases to $2\times 10^{-3}$ at 10~GeV/$c$ and to $10^{-4}$ at 45~GeV/$c$. The momentum-dependence of $P_{\pi e}^+$ is due to the RICH Cherenkov threshold at low momentum, and the similarity of RICH response to $e^+$ and $\pi^+$ at high momentum.

\begin{table}[p]
\caption{Extreme values over the signal momentum range of the $e^\pm$, $\pi^\pm$ identification efficiencies and the $\pi^\pm \rightleftarrows e^\pm$ misidentification probabilities.}
\begin{center}
\vspace{-9mm}
\begin{tabular}{ccrcl}
\hline
Particle type & Identification efficiency & \multicolumn{3}{c}{Misidentification probability}\\
\hline
\rule{0pt}{11pt}$\pi^+$ & \multirow{2}{*}{$98.2\%<\varepsilon_\pi^\pm<98.7\%$} & ~ $10^{-5}$ \!\!\!\!\! & $<~P_{\pi e}^+~<$ & \!\!\!\!\! $2\times 10^{-3}$\\
$\pi^-$ & & ~ 0.8\% \!\!\!\!\! & $<~P_{\pi e}^-~<$ & \!\!\!\!\! 1.1\% \\
\hline
\rule{0pt}{11pt}$e^+$   & $91.0\%<\varepsilon_e^+<96.5\%$ & \multirow{2}{*}{~ 1.2\% \!\!\!\!\!} & \multirow{2}{*}{$<~P_{e\pi}^\pm~<$} & \multirow{2}{*}{\!\!\!\!\! 2.0\%}\\
$e^-$ & $95.5\%<\varepsilon_e^-<97.5\%$ & & \\
\hline
\end{tabular}
\end{center}
\vspace{-9mm}
\label{tab:pid}
\end{table}

\begin{figure}[p]
\begin{center}
\resizebox{0.5\textwidth}{!}{\includegraphics{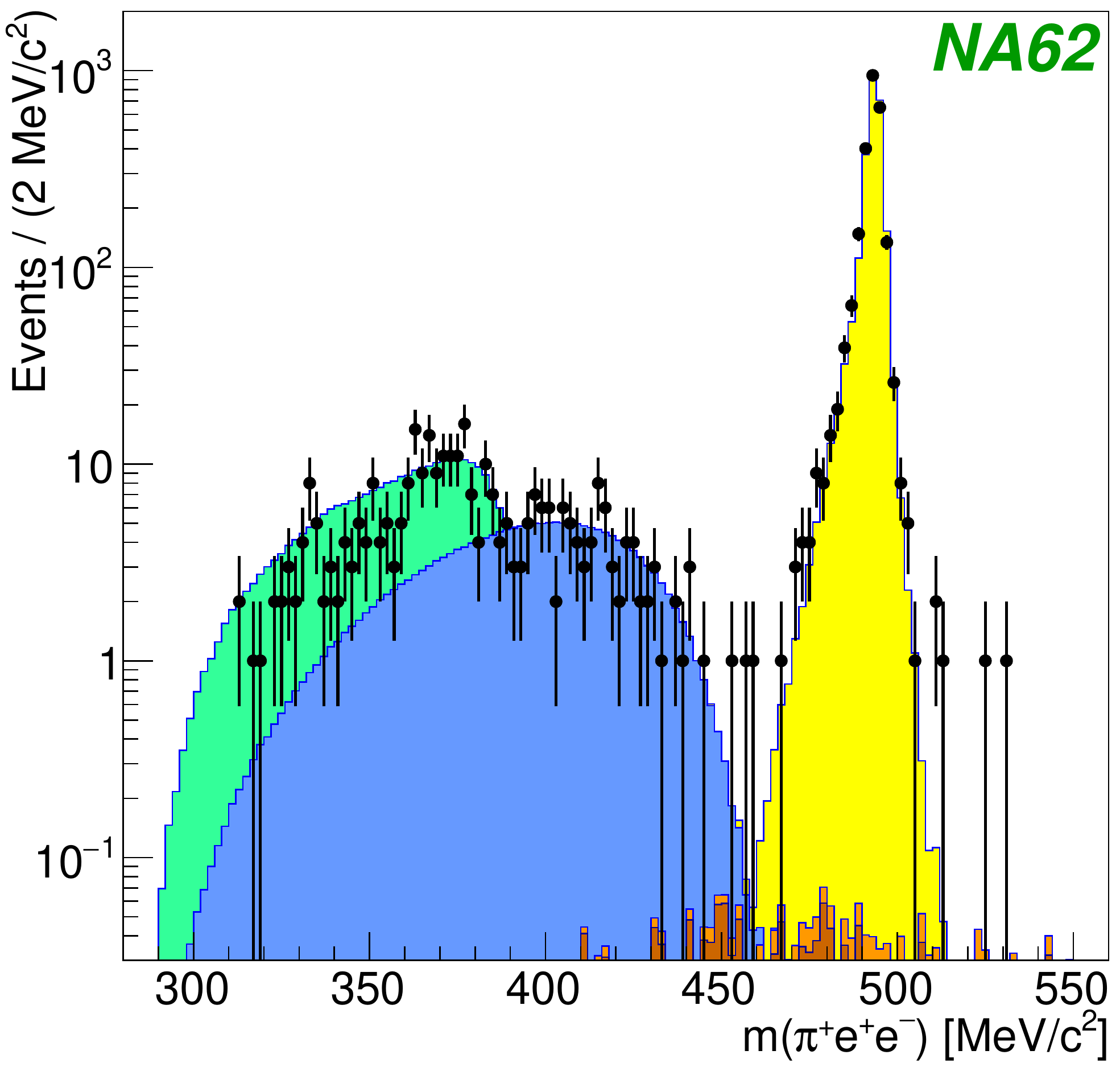}}%
\put(-69,150){$\downarrow$}
\put(-44,150){$\downarrow$}
\resizebox{0.5\textwidth}{!}{\includegraphics{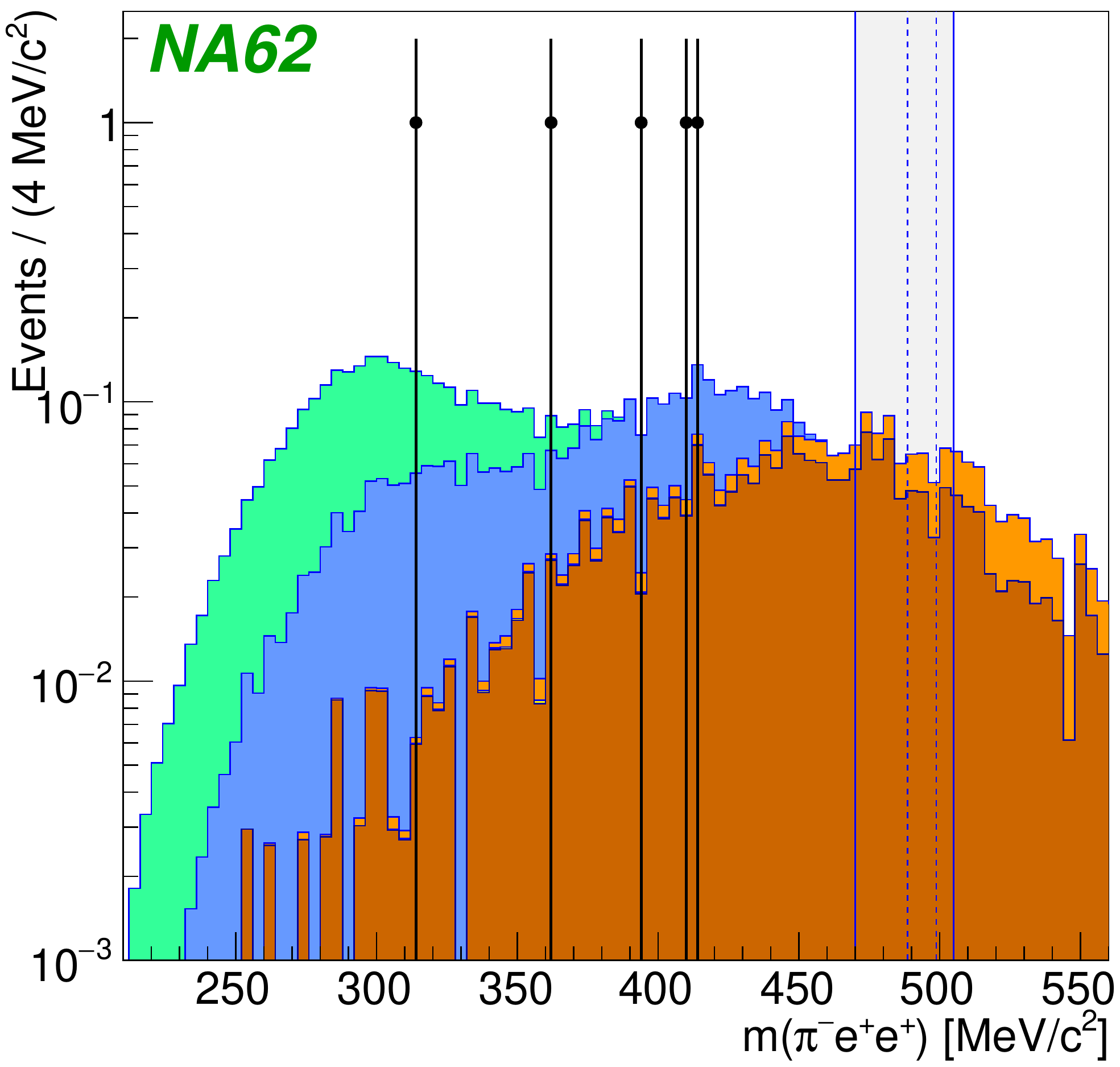}}\\
\resizebox{0.5\textwidth}{!}{\includegraphics{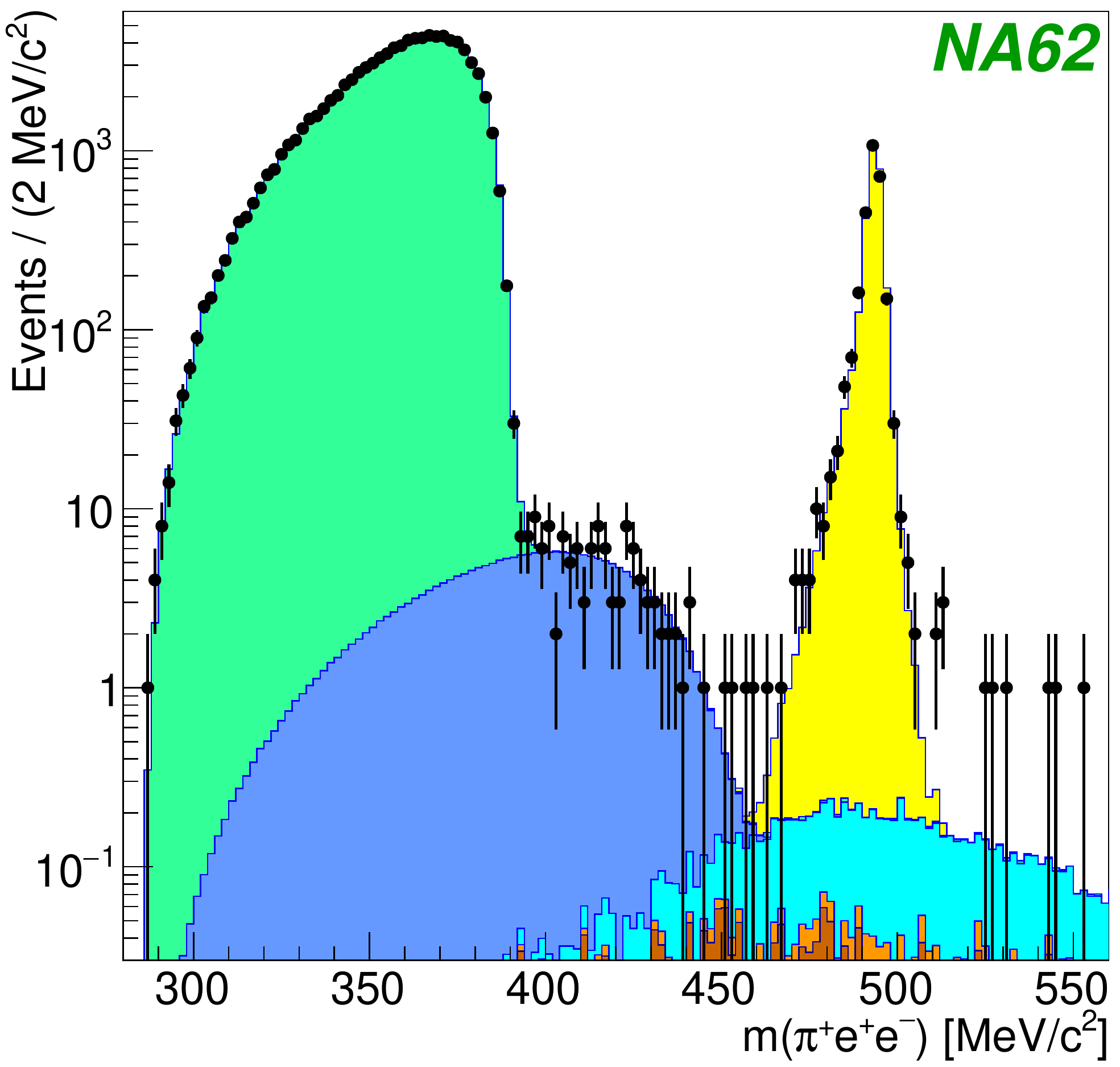}}%
\put(-69,150){$\downarrow$}
\put(-44,150){$\downarrow$}
\resizebox{0.5\textwidth}{!}{\includegraphics{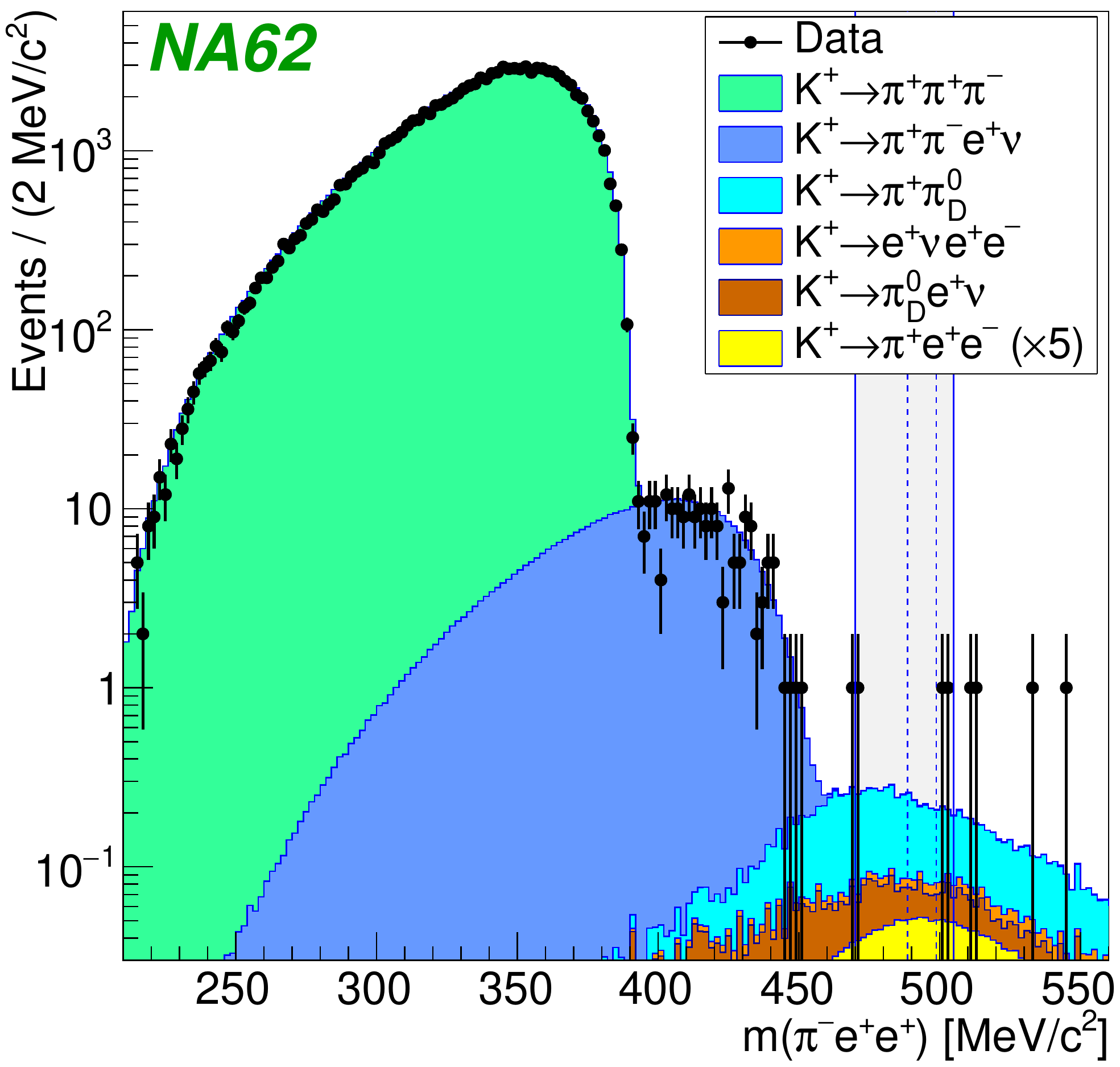}}
\end{center}
\vspace{-14mm}
\caption{Reconstructed mass spectra for the SM (left column) and LNV (right column) $\pi ee$ candidates obtained within the standard selection (top row) and the auxiliary selection without positron identification in the RICH (bottom row). Data are overlayed with background estimates based on simulations. The SM signal region is indicated with arrows. The shaded vertical bands indicate the region masked during the analysis, including the LNV signal region bounded by dashed lines.}
\label{fig:piee}
\end{figure}

The reconstructed $\pi^+e^+e^-$ mass spectra obtained within the standard and auxiliary SM selections, along with the background estimates, are shown in Fig.~\ref{fig:piee} (left). The principal backgrounds in the control mass region are due to $K^+\to\pi^+\pi^+\pi^-$ decays with $\pi^+$ and $\pi^-$ misidentification, and $K^+\to\pi^+\pi^-e^+\nu$ decays with $\pi^-$ misidentification. Positron identification in the RICH reduces the $K^+\to\pi^+\pi^+\pi^-$ background by a factor of 500, with no effect on the $K^+\to\pi^+\pi^-e^+\nu$ background. Contributions involving pion decays in flight $\pi^\pm\to e^\pm\nu$ are found to be negligible. The background in the SM control mass region is simulated to 15\%~(1\%) relative precision within the standard (auxiliary) selection. The limited precision in the former case stems from the dependence of the response of the RICH positron identification algorithm on the event topology in a multi-track environment due to the partial overlap of Cherenkov light-cones, which is difficult to account for accurately.

The reconstructed $\pi^-e^+e^+$ mass spectra obtained within the standard and auxiliary LNV selections are displayed in Fig.~\ref{fig:piee} (right). Due to the presence of two positrons in the LNV final state, backgrounds in the control mass region from $K^+\to\pi^+\pi^+\pi^-$ and $K^+\to\pi^+\pi^-e^+\nu$ decays are reduced by positron identification in the RICH by factors of $5\times 10^4$ and 200, respectively. Five events are observed in the control mass region within the standard selection, in agreement with the expected background from simulation of $5.58\pm0.06_{\rm stat}$. The background in the LNV control mass region within the auxiliary selection is described by simulation to 4\% relative precision. Positron identification in the RICH suppresses the otherwise dominant background to the LNV signal from $K^+\to\pi^+\pi^0_D$ and $K^+\to\pi^+e^+e^-$ decays with $\pi^+$ and $e^-$ misidentification, and reduces the overall estimated background to the LNV signal by a factor of~6. Contributions from $K^+\to\pi^+\pi^0_{DD}$ decays and multiple photon conversions are concluded to be negligible from a study of the data sample selected with vertex charge requirement $q=+3$.

The remaining backgrounds in the LNV signal region are due to $K^+\to\pi^0_D e^+\nu$ and $K^+\to e^+\nu e^+e^-$ decays with $e^-$ misidentified as $\pi^-$. The $K^+\to e^+\nu e^+e^-$ decay is simulated according to Ref.~\cite{bi92}. The contributions from these two decays are estimated to be $0.12\pm0.02_{\rm stat}$ and $0.04\pm0.01_{\rm stat}$ events, respectively. The total expected background in the LNV signal region is
\begin{displaymath}
N_B = 0.16\pm0.03,
\end{displaymath}
where the error includes a systematic uncertainty of 15\% in relative terms to account for the precision of the background description in the control mass regions.

\boldmath
\subsection{$K_{\pi\mu\mu}$ analysis}
\label{sec:pimm}
\unboldmath

Backgrounds to the $K_{\pi\mu\mu}$ processes arise from three-track kaon decays (mainly $K^+\to\pi^+\pi^+\pi^-$) via pion decays in flight and $\pi\rightleftarrows\mu$ misidentification. While the pion decays are implemented accurately in the simulation, misidentification processes cannot be reproduced reliably and require dedicated studies based on control data samples.
\begin{itemize}
\item A pion can be misidentified as a muon due to punch through the iron wall or pileup activity in MUV3. The pileup is simulated using the measured out-of-time signal rates in each MUV3 tile (the mean total signal rate in the MUV3 detector is 16~MHz). The estimated pion to muon misidentification probability, $P_{\pi\mu}(p)$, varies as a function of momentum $p$ from 0.9\% at 5~GeV/$c$ to 0.4\% at 45~GeV/$c$. This dependence arises mainly because the geometrical association of MUV3 signals to tracks involves a search radius whose size varies inversely with momentum to account for multiple scattering.  This optimizes the performance, leading to uniform identification efficiency over momentum and minimal misidentification.
\item A muon can be misidentified as a pion due to MUV3 inefficiency, which is measured using data samples of kinematically selected $K^+\to\mu^+\nu$ decays and beam halo muons to be 0.15\%, with negligible geometric and momentum dependence.
\end{itemize}

The contribution to the LNV sample from $K^+\to\pi^+\pi^+\pi^-$ decays with no pion decays in flight, and both $\pi^+$ misidentified as $\mu^+$, is estimated using a control data sample collected with the multi-track trigger chain (i.e. without muon identification at the trigger level). The full LNV event selection is applied, however the particle identification criteria are inverted to select $\pi^+\pi^+\pi^-$ vertices. Identification of the $\mu^+\mu^+$ pair is then enforced, and a weight of $P_{\pi\mu}(p_1)\cdot P_{\pi\mu}(p_2)\cdot D_1/D_2$ is applied to the event, where $p_{1,2}$ are the reconstructed momenta of the two identified $\pi^+$ tracks, $D_1$ is the downscaling factor of the multi-track trigger chain, and $D_2$ is that of the di-muon chain. The contribution from $K^+\to\pi^+\pi^+\pi^-$ decays with one $\pi^+$ decaying and another $\pi^+$ misidentified as $\mu^+$ is estimated in a similar way using the same data sample, selecting $\pi^+\pi^-\mu^+$ vertices, enforcing identification of the second $\mu^+$ and assigning a weight of $P_{\pi\mu}(p)\cdot D_1/D_2$, where $p$ is the momentum of the identified $\pi^+$ track.

The contribution from $K^+\to\pi^+\pi^+\pi^-$ decay topologies with at least two pion decays in flight, accounting for 70\% of the background in the control mass region, does not necessarily involve pion misidentification and cannot be estimated with the above data-driven method. It is therefore studied with a dedicated simulation. To produce the required MC sample equivalent to ${\cal O}(10^{11})$ $K^+\to\pi^+\pi^+\pi^-$ decays, only the topologies with at least two pion decays in flight (accounting for 4\% of all events) are simulated, and the full simulation of the CHOD, calorimeters and MUV3 is replaced by a fast emulation of their responses. Pion decays in flight typically lead to reconstructed $\pi\mu\mu$ mass values well below the $K^+$ mass. However high mass values within the signal region may be reconstructed due to pion decays in the 7~m long volume between the MNP33 magnet and the third STRAW chamber leading to a biased momentum measurement. The simulation also includes $K^+\to\pi^+\pi^+\pi^-$ decays upstream of the vacuum tank, in which case track bending by the TRIM5 magnet may lead to reconstruction of the decay vertex in the FV with altered kinematic properties.

Contributions to the background from the rare decays $K^+\to\pi^+\mu^+\mu^-$, $K^+\to\pi^+\pi^-\mu^+\nu$, $K^+\to\pi^+\pi^-e^+\nu$, $K^+\to\mu^+\mu^-\mu^+\nu$ are estimated with full simulations. The last process, not measured yet, is simulated according to Ref.~\cite{bi92}. Contributions from the $K^+\to\pi^0_D\mu^+\nu$ and $K^+\to\pi^+\pi^0_D$ decays with ${\cal O}(10^{-3})$ branching fractions and $e^\pm$ particles in the final state are found to be negligible using a technique similar to that described in Section~\ref{sec:piee}. The contribution from multiple in-time kaon decays is found to be negligible using selections with modified track timing consistency requirements, and allowing for multiple vertices.

\begin{figure}[p]
\begin{center}
\resizebox{0.5\textwidth}{!}{\includegraphics{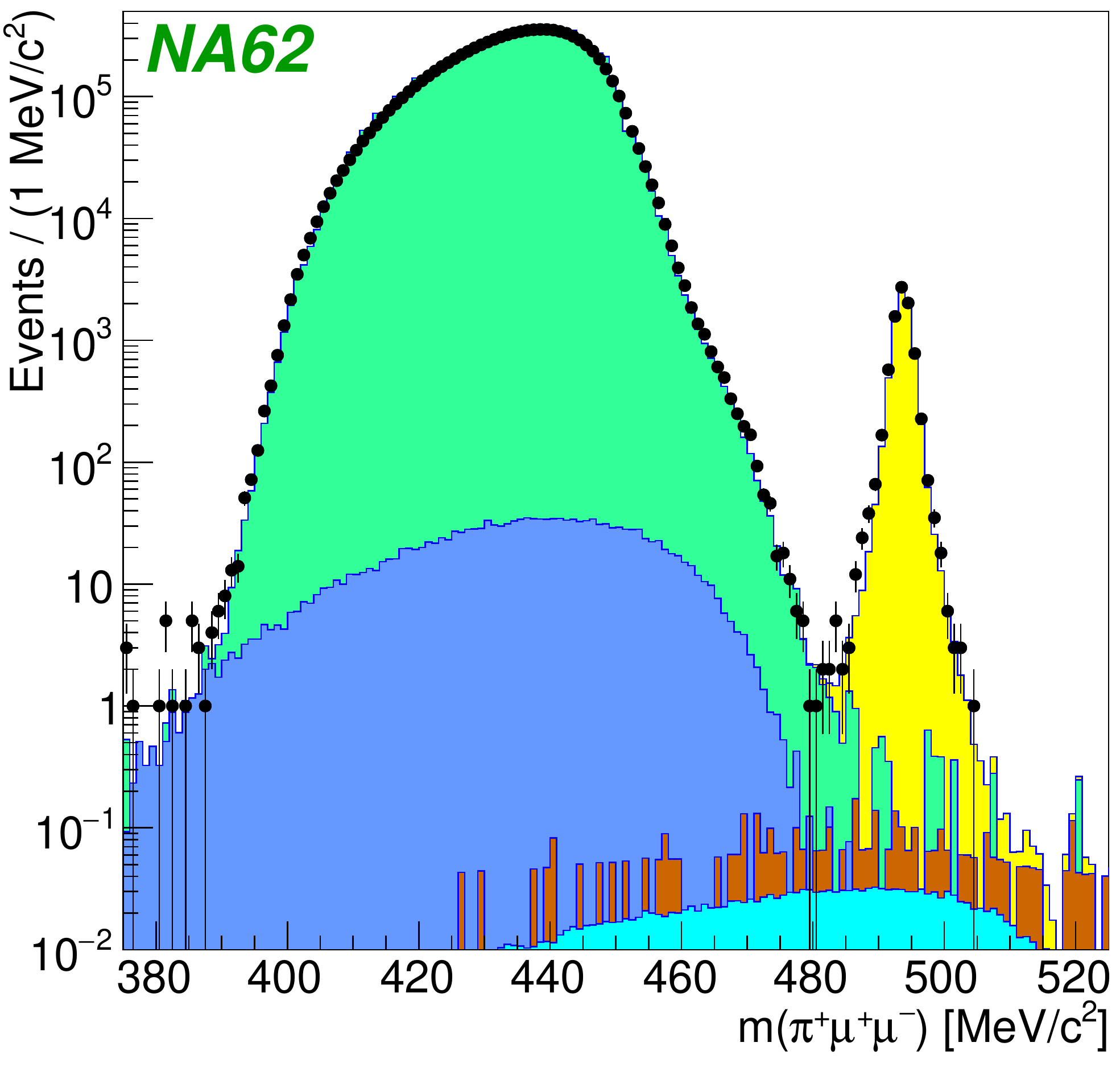}}%
\put(-60,130){$\downarrow$}%
\put(-33,130){$\downarrow$}%
\resizebox{0.5\textwidth}{!}{\includegraphics{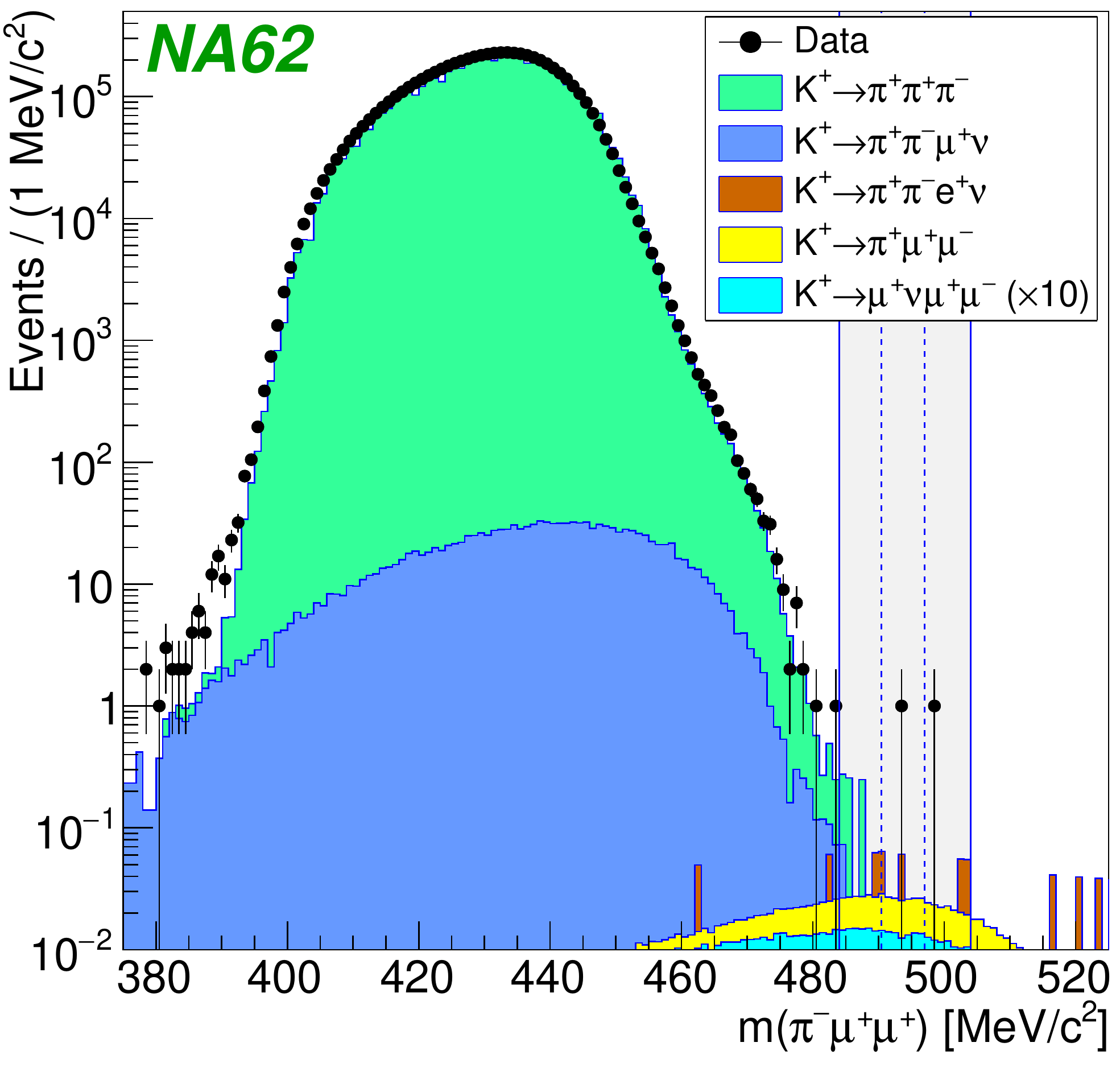}}
\end{center}
\vspace{-14mm}
\caption{Reconstructed mass spectra of the SM $\pi^+\mu^+\mu^-$ (left) and LNV $\pi^-\mu^+\mu^+$ (right) final states: data are overlayed with background estimates based on simulations. Background estimates based on control data samples are not shown. The SM signal region is indicated with arrows. The shaded vertical band indicates the region masked during the analysis, including the LNV signal region bounded by dashed lines.}
\label{fig:pimm}
\end{figure}

The reconstructed $\pi\mu\mu$ mass spectra obtained within the SM and the LNV selections are shown in Fig.~\ref{fig:pimm}. The control-region populations obtained from data and simulation agree to within 3\% for both selections, which validates the background description. The estimated background contributions in the LNV signal mass region from all identified sources are listed in Table~\ref{tab:bkg_pimm}. The expected background is
\begin{displaymath}
N_B = 0.91\pm0.41,
\end{displaymath}
where the uncertainty is statistical due to the sizes of the control and simulated data samples, while the systematic uncertainty is expected to be negligible.

\begin{table}[p]
\caption{Expected backgrounds in the $K^+\to\pi^-\mu^+\mu^+$ signal mass region with their statistical uncertainties.}
\begin{center}
\vspace{-8mm}
\begin{tabular}{lc}
\hline
Process & Expected background \\
\hline
\rule{0pt}{11pt}$K_{3\pi}$ (no $\pi^\pm$ decays) & $0.007\pm0.003$ \\
$K_{3\pi}$ (one $\pi^\pm$ decay) & $0.25\pm0.25$ \\
$K_{3\pi}$ downstream (at least two $\pi^\pm$ decays) & $0.20\pm0.20$ \\
$K_{3\pi}$ upstream (at least two $\pi^\pm$ decays) & $0.24\pm0.24$ \\
$K^+\to\pi^+\mu^+\mu^-$ & $0.08\pm0.02$ \\
$K^+\to\pi^+\pi^-\mu^+\nu$ & $0.05\pm0.05$ \\
$K^+\to\pi^+\pi^-e^+\nu$ & $0.07\pm0.05$ \\
$K^+\to\mu^+\nu\mu^+\mu^-$ & $0.01\pm0.01$ \\
\hline
Total & $0.91\pm0.41$ \\
\hline
\end{tabular}
\end{center}
\vspace{-10mm}
\label{tab:bkg_pimm}
\end{table}


\section{Results}
\label{sec:results}

The information quantifying the sensitivities of the two searches is summarized in Table~\ref{tab:nk}. It includes the numbers of selected SM candidates $N_{\pi\ell\ell}$ used for normalization; the background contaminations (in relative terms) $f_\ell$ in the selected SM decay samples and the acceptances $A_{\pi\ell\ell}$ and $A_{\pi\ell\ell}^{\rm LNV}$ of the SM and LNV selections evaluated with simulation (Section~\ref{sec:bkg}); the branching fractions ${\cal B}_{\pi\ell\ell}$ of the SM decays; the numbers of $K^+$ decays in the FV computed as
\begin{displaymath}
N_K^{\pi\ell\ell}=(1-f_\ell) \cdot N_{\pi\ell\ell}/({\cal B}_{\pi\ell\ell} \cdot A_{\pi\ell\ell});
\end{displaymath}
and the single event sensitivities defined as
\begin{displaymath}
S_{\pi\ell\ell}= \frac{1}{N_K^{\pi\ell\ell}\cdot A_{\pi\ell\ell}^{\rm LNV}} =
\frac{{\cal B}_{\pi\ell\ell}}{(1-f_\ell) \cdot N_{\pi\ell\ell}}\cdot(A_{\pi\ell\ell}/A_{\pi\ell\ell}^{\rm LNV}).
\end{displaymath}

The acceptances are evaluated using the measured phase space densities~\cite{ba09, ba11} for the SM decays, and assuming uniform densities for the LNV decays. The ratios $A_{\pi\ell\ell}/A_{\pi\ell\ell}^{\rm LNV}$ are affected by these assumptions, as well as the charge asymmetry of the geometric acceptance induced by the magnets in the beam line and detector, and also the SM selection condition $m_{ee}>140$~MeV/$c^2$ and positron identification in the RICH in the $K_{\pi ee}$ case. Uncertainties on the ratios $A_{\pi\ell\ell}/A_{\pi\ell\ell}^{\rm LNV}$ are negligible with respect to statistical uncertainties on $N_{\pi\ell\ell}$ and external uncertainties on ${\cal B}_{\pi\ell\ell}$. The ratio $N_K^{\pi\mu\mu}/N_K^{\pi ee}=3.7$ is determined by the downscaling factors of the trigger chains used for the two analyses.

\begin{table}[p]
\caption{Quantities involved in the computation of the single event sensitivities. The most accurate ${\cal B}_{\pi\mu\mu}$ measurement~\cite{ba11} is used rather than the less precise PDG average~\cite{pdg}. The statistical uncertainties on $A_{\pi\ell\ell}$ and $A_{\pi\ell\ell}^{\rm LNV}$ are negligible and the systematic uncertainties, which largely cancel in the acceptance ratios between SM and LNV decays, are not quoted.}
\begin{center}
\vspace{-8mm}
\begin{tabular}{l|cc}
\hline
& $K_{\pi ee}$ analysis & $K_{\pi\mu\mu}$ analysis \\
\hline
SM candidates selected $N_{\pi\ell\ell}$ & 2484 & 8357 \\
Background contamination $f_\ell$ & negligible & $7\times 10^{-4}$ \\
Acceptance $A_{\pi\ell\ell}$ & 3.87\% & 10.93\% \\
Acceptance $A_{\pi\ell\ell}^{\rm LNV}$ & 4.98\% & 9.81\% \\
Branching fraction ${\cal B}_{\pi\ell\ell}\times 10^7$ & $3.00\pm0.09$~\cite{pdg} & $0.962\pm 0.025$~\cite{ba11} \\
Number of decays in FV $N_K^{\pi\ell\ell}/10^{11}$ & $2.14\pm0.04_{\rm stat}\pm0.06_{\rm ext}$ & $7.94\pm0.09_{\rm stat}\pm0.21_{\rm ext}$ \\
Single event sensitivity $S_{\pi\ell\ell}$ & $(0.94\pm 0.03)\times 10^{-10}$ & $(1.28\pm 0.04)\times 10^{-11}$ \\
\hline
\end{tabular}
\end{center}
\vspace{-10mm}
\label{tab:nk}
\end{table}

\newpage

After unmasking the LNV mass regions, no events are observed in the $K_{\pi ee}$ signal region and one event is observed in the $K_{\pi\mu\mu}$ signal region. An additional cross-check of the background estimate is performed in the LNV masked regions but outside the signal regions: no events are (one event is) observed for $K_{\pi ee}$ ($K_{\pi\mu\mu}$), which is consistent with the expectation of $0.46\pm0.04_{\rm stat}$ ($1.05\pm0.46_{\rm stat}$) background events.

Upper limits on the signal branching fractions are obtained using the ${\rm CL_s}$ method~\cite{re02}. In each case, the number of observed events in the LNV signal region and the single event sensitivity with its uncertainty are taken as inputs, and the expected backgrounds are treated using Bayesian inference involving posterior PDFs evaluated assuming uniform prior probabilities.
The resulting upper limits at 90\% CL obtained under the assumption of uniform phase space density are
\begin{eqnarray}
{\cal B}(K^+\to\pi^-e^+e^+) &<& 2.2 \times 10^{-10},\nonumber \\
{\cal B}(K^+\to\pi^-\mu^+\mu^+) &<& 4.2 \times 10^{-11}.\nonumber
\end{eqnarray}
We emphasize that these results, and all other results of searches for LNV decays, depend on  the phase space density assumptions.

\section*{Summary}

Searches for lepton number violating decays $K^+\to\pi^-e^+e^+$ and $K^+\to\pi^-\mu^+\mu^+$ have been performed using about 30\% of the data collected by the NA62 experiment at CERN in 2016--18. The sensitivities are not limited by backgrounds, and the upper limits obtained on the decay rates improve on previously reported measurements by factors of 3 and 2, respectively.


\section*{Acknowledgements}

It is a pleasure to express our appreciation to the staff of the CERN laboratory and the technical
staffs of the participating laboratories and universities for their efforts in the operation of the
experiment and data processing.

The cost of the experiment and its auxiliary systems was supported by the funding agencies of
the Collaboration Institutes. We are particularly indebted to:
F.R.S.-FNRS (Fonds de la Recherche Scientifique - FNRS), Belgium;
BMES (Ministry of Education, Youth and Science), Bulgaria;
NSERC (Natural Sciences and Engineering Research Council), Canada;
NRC (National Research Council) contribution to TRIUMF, Canada;
MEYS (Ministry of Education, Youth and Sports),  Czech Republic;
BMBF (Bundesministerium f\"{u}r Bildung und Forschung) contracts 05H12UM5, 05H15UMCNA and 05H18UMCNA, Germany;
INFN  (Istituto Nazionale di Fisica Nucleare),  Italy;
MIUR (Ministero dell'Istruzione, dell'Universit\`a e della Ricerca),  Italy;
CONACyT  (Consejo Nacional de Ciencia y Tecnolog\'{i}a),  Mexico;
IFA (Institute of Atomic Physics),  Romania;
INR-RAS (Institute for Nuclear Research of the Russian Academy of Sciences), Moscow, Russia;
JINR (Joint Institute for Nuclear Research), Dubna, Russia;
NRC (National Research Center)  ``Kurchatov Institute'' and MESRF (Ministry of Education and Science of the Russian Federation), Russia;
MESRS  (Ministry of Education, Science, Research and Sport), Slovakia;
CERN (European Organization for Nuclear Research), Switzerland;
STFC (Science and Technology Facilities Council), United Kingdom;
NSF (National Science Foundation) Award Number 1506088,   U.S.A.;
ERC (European Research Council)  ``UniversaLepto'' advanced grant 268062, ``KaonLepton'' starting grant 336581, Europe.

Individuals have received support from:
Charles University Research Center (UNCE/SCI/013), Czech Republic;
Ministry of Education, Universities and Research (MIUR  ``Futuro in ricerca 2012''  grant RBFR12JF2Z, Project GAP), Italy;
Russian Foundation for Basic Research  (RFBR grants 18-32-00072, 18-32-00245), Russia;
the Royal Society  (grants UF100308, UF0758946), United Kingdom;
STFC (Rutherford fellowships ST/J00412X/1, ST/M005798/1), United Kingdom;
ERC (grants 268062,  336581 and  starting grant 802836 ``AxScale'').




\end{document}